\documentstyle[prl,aps,multicol,epsfig]{revtex}

\input epsf.tex

\newcommand{\be}{\begin{equation}}
\newcommand{\ee}{\end{equation}}
\newcommand{\ba}{\begin{eqnarray}}
\newcommand{\ea}{\end{eqnarray}}
\newcommand{\ban}{\begin{eqnarray*}}
\newcommand{\ean}{\end{eqnarray*}}

\newcommand{\braket}[2]{\mbox{$ \langle #1 | #2 \rangle $}}

\newcommand{\ket}[1]{\mbox{$ | #1 \rangle $}}
\newcommand{\bra}[1]{\mbox{$ \langle #1 | $}}

\newcommand{\demi}{\frac{1}{2}}

\newcommand{\one}{\leavevmode\hbox{\small1\normalsize\kern-.33em1}}

\newcommand{\moy}[1]{\langle #1 \rangle}

\begin{document}

\title{Direct measurement of superluminal group velocity and of signal velocity \\ in an optical fiber}
\author{Nicolas Brunner, Valerio Scarani, Mark Wegm\"uller, Matthieu Legr\'e and Nicolas Gisin}
\address{Group of Applied Physics, University of Geneva, \\ 20 rue de l'Ecole-de-M\'edecine, CH-1211 Geneva 4, Switzerland}
\date{\today}
\maketitle

\begin{abstract}

We present an easy way of observing superluminal group velocities
using a birefringent optical fiber and other standard devices. In
the theoretical analysis, we show that the optical properties of
the setup can be described using the notion of "weak value". The
experiment shows that the group velocity can indeed exceed $c$ in
the fiber; and we report the first direct observation of the
so-called "signal velocity", the speed at which information
propagates and that cannot exceed $c$.
\end{abstract}

\begin{multicols}{2}

The physics of light propagation is a very timely topic because of
its relevance for both classical \cite{pmdpdl} and quantum
\cite{Qinf} communication. Two kind of velocities are usually
introduced to describe the propagation of a wave in a medium with
dispersion $\omega(k)$: the phase velocity
$v_{ph}=\frac{\omega}{k}$ and the group velocity
$v_{g}=\frac{\partial\omega}{\partial k}$. Both of these
velocities can exceed the speed of light in vacuum $c$ in suitable
cases \cite{jackson}; hence, neither can describe the speed at
which the information carried by a pulse propagates in the medium.
Indeed, since the seminal work of Sommerfeld, extended and
completed by Brillouin \cite{brillouin}, it is known that
information travels at the {\em signal velocity}, defined as the
speed of the front of a square pulse. This velocity cannot exceed
$c$ \cite{forerun}. The fact that no modification of the group
velocity can increase the speed at which information is
transmitted has been directly demonstrated in a recent experiment
\cite{infoFAST}. Superluminal (or even negative) and, on the other
extreme, exceedingly small group velocities, have been observed in
several media \cite{boyd}. In this letter we report observation of
both superluminal and delayed pulse propagation in a tabletop
experiment that involves only a highly birefringent optical fiber
and other standard telecom devices.

Before describing our setup, it is useful to understand in some
more detail the mechanism through which anomalous group velocities
can be obtained. For a light pulse sharply peaked in frequency,
the speed of the center-of-mass is the group velocity $v_g$ of the
medium for the central frequency \cite{jackson}. In the absence of
anomalous light propagation, the local refractive index of the
medium is $n_f$, supposed independent on frequency for the region
of interest. The free propagation simply yields $v_g=L/t_f$ where
$L$ is the length of the medium and $t_f=n_fL/c$ is the free
propagation time. One way to allow fast- and slow-light amounts to
modify the properties of the medium in such a way that it becomes
opaque for all but the fastest (slowest) frequency components. The
center-of-mass of the outgoing pulse appears then at a time
$t=t_f+\moy{t}$, with $\moy{t}$ the mean time of arrival once the
free propagation has been subtracted; obviously $\moy{t}<0$ for
fast-light, $\moy{t}>0$ for slow-light. If the deformation of the
pulse is weak, the group velocity is still the speed of the
center-of-mass, now given by \ba v_g&=&\frac{L}{t_f+\moy{t}}\,.
\label{vgroup} \ea This can become either very large and even
negative ($\moy{t}\rightarrow-\infty$) or very small
($\moy{t}\rightarrow\infty$) --- although in these limiting
situations the pulse is usually strongly distorted, so that our
reasoning breaks down.

\begin{center}
\begin{figure}
\epsfxsize=7cm \epsfbox{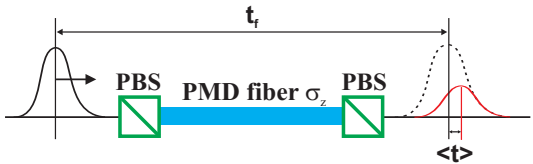} \caption{Basic scheme. By
carefully adjusting the direction of both polarizing beam
splitters (PBS), one can modify the group velocity of the
pulse.}\label{fig1}
\end{figure}
\end{center}

We can now move to our setup, sketched out in Fig. \ref{fig1}. The
medium is a birefringent optical fiber of length $L$, whose
refraction index is $n_f\simeq \frac{3}{2}$ for the telecom
wavelength 1550 nm. The fiber is sandwiched between two
polarizers. This setup is also known in classical optics as a Lyot
filter \cite{lyot}. As shown in \cite{prl}, this simple situation
can be described with the quantum formalism of weak values
\cite{weak}. In fact, the three essential ingredients of weak
values are present here: (I) The input polarizer allows the {\em
pre-selection} of a pure polarization state $\ket{\psi_0}$. (II)
The birefringent fiber performs a {\em pre-measurement} of
polarization by spatially separating the two fiber's eigenmodes,
supposed to be described by the eigenstates $\ket{H}$ and
$\ket{V}$ of $\sigma_{z}$ --- that is, we choose $z$ to be the
birefringence axis. This pre-measurement, known in telecom physics
as polarization-mode dispersion (PMD), is weak whenever the
temporal shift $\delta\tau$ (the so-called differential group
delay, DGD) between the two eigenmodes of the fiber is much
smaller than the coherence time $t_{c}$ of the optical pulse.
(III) The output polarizer performs the {\em post-selection} on
another chosen polarization state $\ket{\phi}$. The mean time of
arrival of the optical pulse can be expressed as a function of a
weak value $W$ \cite{prl,pmdpdl}: \ba \label{main} \moy{t} &=
&\frac{\delta\tau}{2}\,\mbox{Re}W\,=\,\frac{\delta\tau}{2}
\textrm{Re} \bigg[
\frac{\bra{\phi}\sigma_{z}\ket{\psi}}{\braket{\phi}{\psi}} \bigg]
\label{weak1}\ea where $\ket{\psi}$ is the polarization obtained
by the free rotation of $\ket{\psi_0}$ in the fiber. Thus,
$\moy{t}$ is in principle unbounded: by carefully choosing
$\ket{\psi_0}$ and $\ket{\phi}$, i.e the direction of each
polarizer, one can tune the group velocity (\ref{vgroup}) to any
desired value. In summary, by appending two polarizers at the ends
of a birefringent fiber, one creates an {\em effective medium}
whose dispersion and absorption properties depend on the pre- and
post-selection; in this effective medium, the group velocity can
be increased or decreased by an arbitrary amount by merely
choosing the orientations of two polarizers.

The rest of the paper is organized in three parts: first we derive
the optical properties (dispersion and absorption) of our
effective medium, stressing the role of the weak value $W$. Then
we describe the experiment and present our main results in Figs
\ref{4graphs} and \ref{fasterthanlight}. We shall conclude by
comparing our experiment to a related one \cite{solli91}, and by
stressing the differences with other fast- and slow-light
techniques.

{\em Theoretical analysis.} For any linear medium in which light
propagates, the absorption coefficient $\kappa (\omega)$ and the
index of refraction $n(\omega)$ are defined through \ba
\label{gdef} G(\omega)&= &e^{-\kappa (\omega) L }\, e^{i
\frac{n(\omega)\omega}{c}L} \ea where $L$ is the length of the
fiber, and where $G(\omega)$ is the linear response function which
characterizes the evolution of a plane wave of frequency $\omega$
in the medium: $e^{-i\omega t} \rightarrow G(\omega)e^{-i\omega
t}$. The calculation of $G(\omega)$ for the effective medium
sketched in Fig. \ref{fig1} has to take polarization into account.
The first polarizer prepares the state \ba \ket{\psi_0}
\,\otimes\, e^{-i\omega t}&=&(a_0 \ket{H}+ b_0 \ket{V}) \otimes
e^{-i\omega t} \ea where $\ket{H}$ and $\ket{V}$ are the
eigenmodes of the fiber, the eigenstates of the Pauli matrix
$\sigma_{z}$ for the eigenvalues $\pm 1$. As discussed in Ref.
\cite{prl}, the evolution of the polarization of the plane wave
because of birefringence is given by the unitary operator
$e^{i\omega\delta\tau\, \sigma_{z}/2}$, which describes a global
rotation around the $z$ axis of the Poincar\'e sphere --- with our
conventions, $\ket{H}$ is the slow mode and $\ket{V}$ is the fast
mode. So, the state that reaches the second polarizer is
$\ket{\psi(\omega)}\,\otimes\,e^{in_{f}\omega L/c}\,e^{-i\omega
t}$ where one recognizes the phase acquired during free
propagation through the fiber and where \ba
\ket{\psi(\omega)}&=&a_0\,
e^{i\frac{\delta\tau}{2}\omega}\ket{H}\,+\, b_0\,
e^{-i\frac{\delta\tau}{2}\omega}\ket{V}\,. \label{statepsi}\ea The
second polarizer is represented by the projector on a polarization
state $\ket{\phi}=a_1 \ket{H}+ b_1 \ket{V}$. So the output state
reads $\ket{\phi} \,\otimes\, G(\omega)e^{-i\omega t}$ with a
response function \cite{note1} \ba G(\omega) 
&=& e^{in_{f}\omega L/c}\,(A+B)\,{\cal F}(\omega,W_0)
\label{gpmd}\ea where we have written $A\equiv a_1^{*}a_0$ and
$B\equiv b_1^{*}b_0$, ${\cal F}(\omega,W_0) =
\cos(\omega\,\delta\tau/2)+ i W_0 \sin(\omega\,\delta\tau/2) $ and
where \ba W_0&=&\frac{A-B}{A+B} \,=\, \frac{\bra{\phi}
\sigma_{z}\ket{\psi_0}}{\braket{\phi}{\psi_0}}
\,\equiv\,W_R+iW_I\ea is the weak value involving the orientations
of the two polarizers \cite{solli92}. Note that this is not the
weak value that enters the mean time of arrival (\ref{weak1}):
that one is computed using $\ket{\psi(\omega)}$ instead of
$\ket{\psi_0}$; using (\ref{statepsi}), one obtains \ba W&=&
\frac{W_R+i\left(W_I\cos \omega\delta\tau+\demi(1-|W_0|^2) \sin
\omega\delta\tau\right)}{\left|{\cal F}(\omega,W_0)\right|^2}
\label{weak2}\,. \ea From (\ref{gdef}) and (\ref{gpmd}) it is
straightforward to derive expressions for the absorption
coefficient $\kappa (\omega)$ and the index of refraction
$n(\omega)$. For simplicity, the global phases of both
polarization states $\ket{\psi_0}$ and $\ket{\phi}$ are chosen
such that $(A+B)$ is real and contributes thus only to the
absorption. The results are: \ba \label{netc} \kappa
(\omega)&=&-\frac{1}{L} \left[ \ln(A+B)+
\ln\left|{\cal F}(\omega,W_0)\right| \right]\\
\label{nph} n(\omega)&=&n_{f}+\frac{c}{L \omega} \arctan\left[
\frac{W_R}{\cot(\omega\delta\tau/2)-W_I}\right]\,. \ea From
$n(\omega)$, one can derive the group index
$n_{g}(\omega)=n+\omega \frac{dn}{d \omega}$. Using (\ref{weak2})
and (\ref{weak1}) one finds \ba \label{ng1} n_{g}(\omega)&=&
n_{f}+c\,\frac{\delta\tau}{2L}\,
\mbox{Re}W\,=\,n_f\,+\,\frac{c}{L}\moy{t}\,. \label{ng}\ea This
gives exactly the behavior (\ref{vgroup}) for the group velocity
--- indeed, both derivations are based on the assumption that the
coherence time of the pulse is large compared to $\delta\tau$.
When $\mbox{Re}W=\frac{W_{R}}{\left|{\cal
F}(\omega,W_0)\right|^2}$ is small, the second term in (\ref{ng})
disappears and $n_{g}\approx n_{f}$: this is normal propagation,
obtained for most choices of the pre- and post-selection. The
conditions $\mbox{Re}W=\pm 1$ are obtained for the slow, resp. the
fast, polarization mode of the fiber, $\ket{\psi_0}=\ket{H}$, resp
$\ket{V}$. When $\mbox{Re}W>1$, we obtain the slow-light regime;
when $\mbox{Re}W<-1$, the fast-light regime. In particular, since
$n_f\simeq\frac{3}{2}$, superluminal group velocity ($n_g<1$)
requires $\mbox{Re}W<-\frac{L}{c\delta\tau}$, negative group
velocity ($n_g<0$) requires $\mbox{Re}W<-3\frac{L}{c\delta\tau}$.

\begin{center}
\begin{figure}
\epsfxsize=7cm \epsfbox{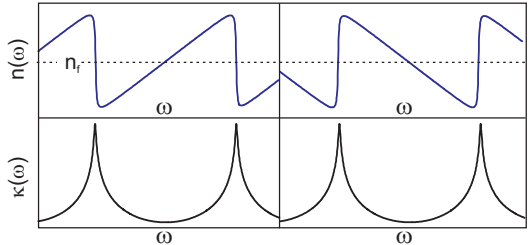} \caption{Index of refraction
and absorption coefficient versus frequency for $W_0=W_R=\pm 60$.}
\label{refrabs}\end{figure}
\end{center}

In fig. \ref{refrabs}, the absorption coefficient (\ref{netc}) and
the index of refraction (\ref{nph}) are plotted for different weak
values. When $W_{R}$ increases, the slope of the index of
refraction becomes steeper: a larger positive (negative) group
delay corresponds to a slower (faster) group velocity. It is also
clear from fig. \ref{refrabs} that a strong absorption occurs
whenever the change in the group velocity is important. Although
these features are usual in their qualitative formulation, we note
(as the authors of Ref. \cite{solli91} did) that the standard
Kramers-Kronig relations do not apply: these relations predict
that the absorption coefficient ("imaginary part of the index")
determines uniquely the index of refraction ("real part"), and
viceversa; while in our case, the same absorption is associated to
different dispersions. Modified Kramers-Kronig relations have been
discussed in \cite{toll}. However, a full study of the question
lies beyond the approach presented here, because our formula
(\ref{gpmd}) for $G(\omega)$ is valid only in a limited frequency
range --- e.g., the index of the fiber is not constant over all
possible frequencies.

\begin{center}
\begin{figure}
\epsfxsize=8cm \epsfbox{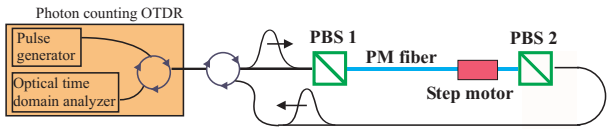}
  \caption{Experimental setup} \label{figexp}
\end{figure}
\end{center}
{\em Experiment}. The experimental setup is sketched out in fig.
\ref{figexp}. As a source and for detection we use an optical time
domain reflectometer (OTDR). This is a telecom instrument designed
to measure loss profiles of fibers: it sends short laser pulses
and analyzes the amount of back-scattered light as a function of
time. Here we use a commercial prototype OTDR working in photon
counting mode at the telecom wavelength $\lambda=1.55\, \mu$m
\cite{otdr}. It contains a pulsed DFB laser ($\Delta \nu\simeq
500$MHz , pulse duration 2 ns) and a gated peltier-cooled InGaAs
photon counter (gate duration 2 ns). The photon counting OTDR is
well adapted to this experiment since it allows to monitor the
optical pulse, even in the presence of strong absorption. The
birefringent fiber is a polarization maintaining (PM) fiber of
length $L=1.5$ m, with a DGD $\delta\tau=2.66$ ps, measured with
an interferometric low-coherence method \cite{mesdgd}. The fiber
is placed between two polarizing beam splitter (PBS) cubes with
specified extinction ratios of 50 dB. Both cubes are mounted on
rotational stages, allowing a precise alignment of pre- and
post-selected polarization states. Furthermore a micrometer step
motor permits to slightly change the length of the fiber. This
allows alignment of the post-selection also in the $x-y$ plane of
the Poincar\'e sphere, since polarization is rotating around the
$z$ axis in the fiber.

The crucial part of the experiment is the alignment of both PBS
cubes. The input polarizer has to be aligned at $45^{o}$ relative
to the fiber's axis \cite{45}. This is done by injecting
incoherent light (from a LED) into the system and minimizing the
degree of polarization at the output of the PM fiber. For the
post-selection, the position of the second PBS cube and the length
of the fiber are adjusted such that transmission through the
system is minimum.

A first set of results is presented in fig. \ref{4graphs}. Figure
(a) shows the data of a sequence of successive measurements on a
dB scale. The largest curve is the reference pulse, i.e. the pulse
in the normal regime, without any specific alignment of the PBS
cubes. Between each measurement the post-selection is slightly
changed, in order to decrease transmission: we observe that the
lower the transmission, the higher the group velocity, in
agreement with the theory. In addition, figure (a) clearly
illustrates the difference between group and signal velocities
\cite{brillouin}. In fact, even though the group velocity is
higher for each successive curve, the signal velocity remains
constant and equal to $c/n_f$, since the front parts of all pulses
are strictly identical. Figures (b) and (c) show, respectively, a
fast- and a slow-light example. The measured data are plotted
together with the reference pulse (dashed curve) and a fit (smooth
solid curve). The fit is obtained by applying the following
transformation to the observed intensity of the reference pulse
$I_{in}(t)$: $I_{in}(t)\rightarrow A_{in}(t)=\sqrt{I_{in}(t)}
\stackrel{FFT}{\rightarrow }A_{in}(\omega) \rightarrow
A_{out}(\omega)=G(\omega)A_{in}(\omega) \stackrel{FFT}{\rightarrow
}A_{out}(t)\rightarrow I_{out}(t)=|A_{out}(t)|^2$, where FFT is a
discrete Fourier transform procedure. The first step of this
fitting procedure assumes the reference pulse to be
Fourier-transform limited; the slight discrepancies found between
the data and the fits (in the front and back parts of the pulses)
are probably due to this approximation. The fitting parameter is
the weak value $W$, supposed to be real. Note that in (b), $|W|$
lies below $\frac{L}{c\delta\tau}\simeq 2000$, the value needed in
order to demonstrate superluminal group velocity. To reach that
bound, much longer input pulses are needed. We replace the OTDR
source by an external source: a DFB laser ($\Delta\nu\simeq 2$
MHz) working in CW mode, modulated by an external electro optic
modulator (EOM). This source creates nearly gaussian pulses with a
coherence time of about 50ns. The OTDR is still used for detection
and triggers the modulator. Results are presented in fig.
\ref{fasterthanlight}. The pulse on the right shows clearly
superluminal group velocity. By fitting the position of the
maximum of the output pulse \cite{note3}, we find $W\simeq -3500$,
consistent with superluminal but non-negative group velocity, as
expected.

{\em Comparison with other experiments.} Solli and co-workers
reported on a similar, but indirect, experiment using polarized
microwaves and a photonic crystal as birefringent medium, with
identical conclusions on the group velocity \cite{solli91}. Their
conclusions were extracted from phase measurements only, so in
particular they had no result on the signal velocity.

Most experiments on slow light have been performed using
electromagnetically induced transparency (EIT) \cite{boyd,EIT}. In
EIT, the mode that carries information can be transmitted without
losses, which is not the case in our experiment.


In conclusion, we presented an easy way to create slow and fast
light using a birefringent optical fiber and other standard
telecom components. The theoretical part of this work was mainly
devoted to the study of the linear response function of our system
$G(\omega)$, that is determined by a weak value, thus extending
the previous work of \cite{prl}. In the experiment, we obtained
clear evidence for superluminal group velocities (fig.
\ref{fasterthanlight}); we also gave the first direct measurement
of the "signal velocity" [fig. \ref{4graphs} (a)], showing that
the increase of the group velocity does not increase the speed at
which information travels.

We thank C. Barreiro for technical support. This work was
supported by the Swiss NCCR "Quantum photonics".

\begin{center}
\begin{figure}
\epsfxsize=7cm \epsfbox{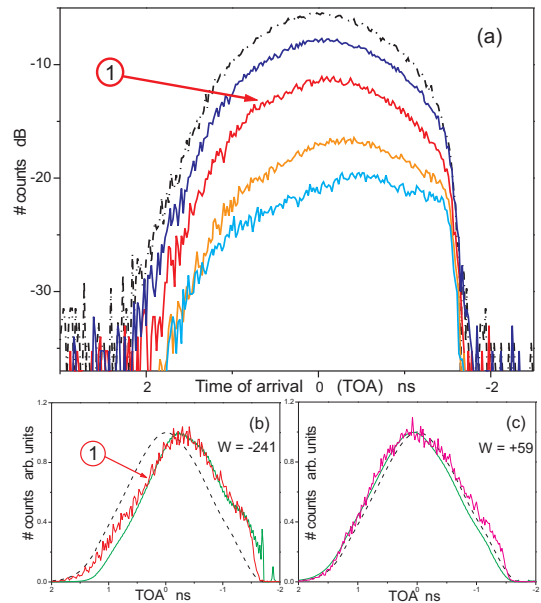} \caption{In the three
figures, the reference pulse (dashed) has been normalized for
convenience. (a) Sequence of measurements obtained by varying the
orientation of a polarizer, in a log scale. The stronger the
absorption, the larger the group velocity, as expected. However,
the pulse is distorted in such a way that its front travels with a
constant speed, the signal velocity $c/n_f$. (b),(c) Examples of
measurements of fast-light and of slow-light, in a linear scale,
together with theoretical fits (smooth solid curve). The real weak
value $W$ obtained from the fit is given.} \label{4graphs}
\end{figure}
\end{center}

\begin{center}
\begin{figure}
\epsfxsize=7cm \epsfbox{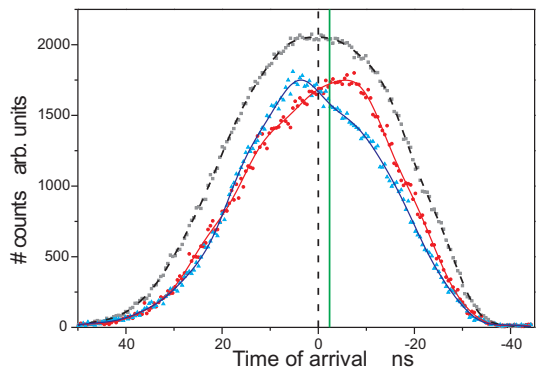} \caption{Fast- and
slow-light curves compared to the reference pulse, whose center is
marked by the dashed vertical line. The full vertical line marks
the position of the summit of a pulse that would have travelled at
speed $c$. The fast pulse shows superluminal group velocity.}
\label{fasterthanlight}
\end{figure}
\end{center}

\end{multicols}


\begin{thebibliography}{99}

\bibitem{pmdpdl}
B. Huttner, C. Geiser and N. Gisin, IEEE J. Sel. Top. Quantum
Electron. {\bf 6}, 317 (2000).

\bibitem{Qinf} D. Bouwmeester, A. Ekert and A. Zeilinger (eds), {\em The physics of quantum
information} (Springer, Berlin, 2000).

\bibitem{jackson} J. D. Jackson, {\em Classical Electrodynamics}, 2nd edition (Wiley, New-York,
1975), Sect. 7-8.

\bibitem{brillouin} L. Brillouin, {\em Wave propagation and group velocity} (Academic
Press, New York, 1960).

\bibitem{forerun} Actually, some
weak excitations called "forerunners" may arrive even before the
main front, but not faster than c. Their speed has been called,
somehow unfortunately, "front velocity".

\bibitem{infoFAST} M. D. Stenner, D.J. Gauthier and M. Neifeld, Nature {\bf 425},
695 (2003)

\bibitem{boyd} R. W.  Boyd, D. J. Gauthier, Progress in Optics Vol. 43,
edited by E. Wolf (Elsevier, Amsterdam, 2002)

\bibitem{lyot} W. Demtr\"oder, {\em Laser spectroscopy} (Springer, Berlin, 2003)

\bibitem{prl} N. Brunner, A. Ac\'{\i}n, D. Collins, N. Gisin, V. Scarani, Phys. Rev. Lett.  {\bf 91}, 180402 (2003)

\bibitem{weak} For a review see: Y. Aharonov and L. Vaidman,
quant-ph/0105101; published in: J.G. Muga, R. Sala Mayato, I. L.
Egusquiza (eds), {\em Time in Quantum Mechanics}, Lecture Notes in
Physics (Springer, Berlin, 2002).

\bibitem{solli91} D. R. Solli, C.F. McCormick, C. Ropers, J.J. Morehead, R.Y. Chiao,
J.M. Hickmann, Phys. Rev. Lett. {\bf 91}, 143906 (2003).



\bibitem{note1} We have found a scalar response function
$G(\omega)$ because we have pre- and post-selected on pure
polarization states. In general, the response function
$G_{ij}(\omega)$ will be tensorial, relating the polarization
components \cite{solli91}.

\bibitem{solli92} A more complex link between response functions and weak values in a similar situation was
pointed out in: D. R. Solli, C.F. McCormick, R.Y. Chiao, S.
Popescu, J.M. Hickmann, Phys. Rev. Lett. {\bf 92}, 043601 (2004).

\bibitem{toll} J.S. Toll, Phys. Rev. {\bf 104}, 1760 (1956)

\bibitem{otdr} M. Wegm\"uller, F. Scholder, N. Gisin, J. Lightwave Tech. {\bf 22}, 390 (2004)

\bibitem{mesdgd} N. Gisin, J.P. Pellaux, J.P. von der Weid, IEEE
J. Lightwave Technology {\bf 9}, 821 (1991)

\bibitem{45} It can be easily shown from (\ref{main}) that this pre-selection
maximizes the mean time of arrival $\moy{t}$.

\bibitem{note3} We were not able to correctly fit
the shape of these curves. This is probably due to a saturation of
the OTDR detection or to some phase noise that invalidates the
first step of the fitting procedure.

\bibitem{EIT} S. Harris, Physics Today, July 1997; M.D. Lukin, A.
Imamo\u{g}lu, Nature {\bf 413}, 273 (2001).





\end{thebibliography}
\end{document}